# Experimental realization of a primary-filling e/3 quasiparticle interferometer

F. E. Camino, Wei Zhou, and V. J. Goldman

*Department of Physics, Stony Brook University, Stony Brook, NY 11794-3800, USA*

We report experiments on a Laughlin quasiparticle interferometer where the entire system is on the 1/3 primary fractional quantum Hall plateau. Electron-beam lithography is used to define an approximately circular 2D electron island separated from the 2D bulk by two wide constrictions. The interferometer consists of counterpropagating chiral edge channels coupled by quantum-coherent tunneling in the two constrictions, thus enclosing an island area. Interference fringes are observed as conductance oscillations, similar to the Aharonov-Bohm effect. The flux and charge periods of the interferometer device are calibrated with electrons in the integer quantum Hall regime. In the fractional regime we observe magnetic flux and charge periods $h/e$ and $e/3$, respectively, corresponding to creation of one quasielectron in the island. Quantum theory predicts a $3h/e$ flux period for charge $e/3$, integer statistics particles. Accordingly, the observed periods demonstrate the anyonic statistics of Laughlin quasiparticles.

## I. INTRODUCTION

A clean system of 2D electrons subjected to high magnetic field at low temperatures condenses into the fractional quantum Hall (FQH) fluids.[1-4] An exact filling $f$ FQH condensate is incompressible and gapped, the celebrated examples of FQH condensates are the Laughlin many-electron wave functions for the *primary* fillings $f = 1/(2j+1)$, with $j$ an integer. The elementary charged excitations of an FQH condensate are the Laughlin quasiparticles. Deviation of the filling factor from the exact value is achieved by excitation of either quasielectrons or quasiholes out of the condensate; at such fillings the ground state of an FQH fluid consists of the quasiparticle-containing condensate. The FQH quasiparticles have fractional electric charge[2-6] and obey fractional statistics.[7-10]

Fractionally charged quasiparticles were first observed directly in quantum antidot experiments, where quasiperiodic resonant conductance peaks are observed when the occupation of the antidot is incremented by one quasiparticle.[6,11,12] A quantum antidot is a small potential hill, defined lithographically in the 2D electron system. Complementary geometry where a 2D electron island is defined by two nearly open constrictions comprises an electron interferometer.[13-16] Recently, we reported realization of a quasiparticle interferometer where $e/3$ quasiparticles of the $f = 1/3$ FQH fluid execute a closed path around an island of the $f = 2/5$ fluid.[9,10,17] The interference fringes were observed as conductance oscillations as a function of the magnetic flux through the island, that is, the Aharonov-Bohm effect. The observed flux and charge periods, $\Delta_\Phi = 5h/e$ and $\Delta_Q = 2e$, are equivalent to excitation of ten $q = e/5$ quasiparticles of the 2/5 fluid. Such superperiodic $\Delta_\Phi > h/e$ had never been reported before in any system. The superperiod is interpreted as imposed by the topological order of the underlying FQH condensates,[18] manifested by the anyonic statistical interaction of the quasiparticles.[19,20]

Our present experiment utilizes a comparable quasiparticle interferometer, but with much less depleted constrictions, Fig. 1. This results in the entire electron island being at the primary quantum Hall filling $f = 1/3$ under coherent tunneling conditions. In this regime the interfering $e/3$ quasiparticles execute a closed path around an island of the 1/3 FQH fluid containing other $e/3$ quasiparticles. For the first time in such devices we report interferometric oscillations. The flux and charge periods of $\Delta_\Phi = h/e$ and $\Delta_Q = e/3$, respectively, correspond to addition of one quasiparticle to the area enclosed by the interference path. These periods are the same as in quantum antidots, but the quasiparticle path encloses no electron vacuum in the interferometer. The results are consistent with the Berry phase quantization condition that includes both Aharonov-Bohm and anyonic statistical contributions. This simpler regime of one quantum Hall filling only should facilitate theoretical consideration of the quasiparticle interferometer physics.

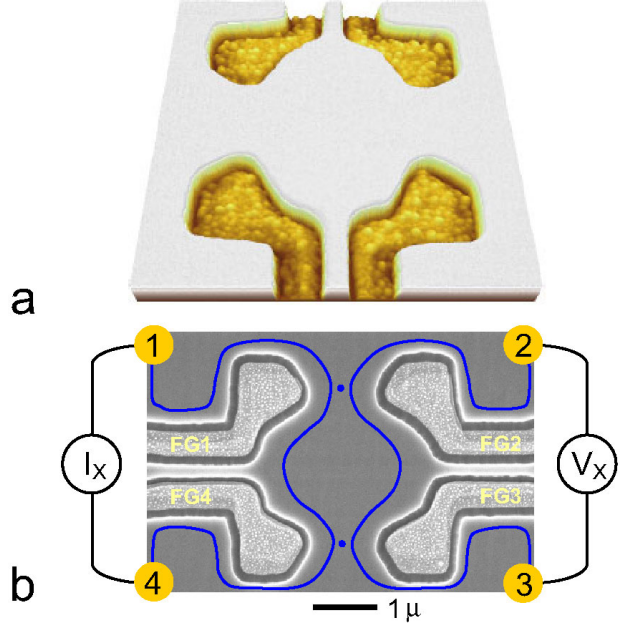

FIG. 1. The primary filling quasiparticle interferometer device. Atomic force (a) and scanning electron (b) micrographs. Four front-gates (FG) are deposited in shallow etch trenches. Depletion potential of the trenches defines the electron island. The chiral edge channels (blue lines) follow equipotentials at the periphery of the undepleted 2D electrons. Tunneling (blue dots) occurs at the saddle points in the two constrictions. The edge channel path is closed by the tunneling links, thus forming the interferometer. The backgate (not shown) extends over the entire 4×4 mm sample.

## II. EXPERIMENTAL RESULTS

The interferometer devices were fabricated from low disorder AlGaAs/GaAs heterojunctions. The four independently-contacted front gates were defined by electron beam lithography on a pre-etched mesa with Ohmic contacts. After a shallow 130 nm wet etching, Au/Ti front-gate metal was deposited in the etch trenches, followed by lift-off, Fig. 1(a,b). Even when front gate $V_{FG} = 0$, the GaAs surface depletion of the etch trenches creates electron confining potential defining two wide constrictions, which separate an approximately circular electron island from the 2D "bulk".

Samples were mounted on sapphire substrates with In metal, which serves as the global backgate, and were cooled in a dilution refrigerator to 10.2 mK bath temperature, calibrated by nuclear orientation thermometry. Extensive cold filtering cuts the electromagnetic environment incident on the sample, allowing to achieve electron temperature $\leq 15$ mK in an interferometer device.[21] Four-terminal resistance $R_{XX} = V_X / I_X$ was measured with 50 pA ($f = 1/3$) or 200 pA ($f = 1$), 5.4 Hz AC current injected at contacts 1 and 4. The resulting voltage $V_X$, including the interference signal, was detected at contacts 2 and 3. The Hall resistance $R_{XY} = V_{4-2}/I_{3-1}$ is determined by the quantum Hall filling $f$ in the constrictions, giving definitive values of $f$. The oscillatory conductance $\delta G$ is calculated from the directly measured $\delta R_{XX}$ and the quantized Hall resistance $R_{XY} = 3h/e^2$ as $\delta G = \delta R_{XX}/(R_{XY}^2 - \delta R_{XX} R_{XY})$, a good approximation for $\delta R_{XX} << R_{XY}$.

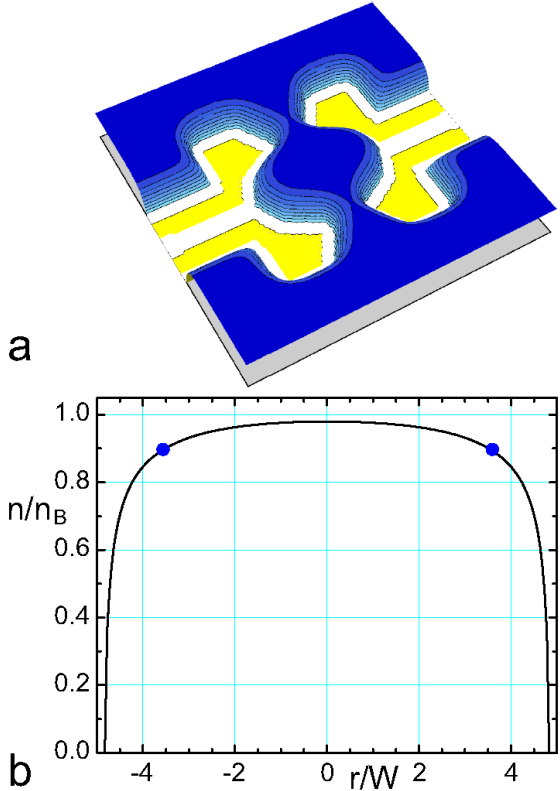

a

b

FIG. 2. (a) Illustration of the 2D electron density profile of the primary-filling interferometer device. Conductance oscillations are observed when the counterpropagating edge channels are coupled by tunneling at the saddle points in the constrictions. (b) Calculated radial profile of the electron density in a circular island defined by an etched annulus of inner radius 1340 nm and bulk density $n_B = 1.25 \times 10^{11}$ cm$^{-2}$. The $B = 0$ island depletion model is described in Ref. 14; depletion length parameter $W = 230$ nm. The blue circles give the edge ring radius $r = 825$ nm, obtained from the experimental Aharonov-Bohm period. The edge ring density is then 0.92 of the island center density, and the entire interferometer has the same filling for wide quantum Hall plateaus.

The etch trenches define two 1.25 µm wide constrictions, which separate an approximately circular electron island from the 2D bulk. Moderate front-gate voltages $V_{FG}$ are used to fine tune the constrictions for symmetry of the tunnel coupling and to increase the oscillatory interference signal. The $B = 0$ shape of the electron density profile is predominantly determined by the etch trench depletion, illustrated in Fig. 2. For the 2D bulk density $n_B = 1.25 \times 10^{11}$ cm$^{-2}$ there are ~3,500 electrons in the island. The depletion potential has saddle points in the constrictions, and so has the resulting density profile.

In a quantizing magnetic field, the overall electron density profile follows the $B = 0$ profile in order to minimize the large Coulomb charging energy arising from significant deviations from the donor-neutralizing $B = 0$ profile. This results in the deviation of the local electron density from that corresponding to the exact, incompressible quantum Hall filling. The density deviation is accommodated by "excitation" of electrons to the next partially occupied Landau level, or holes in the highest fully occupied level. The excitation costs a fraction of the quantum Hall gap energy, but saves the macroscopic charging energy. The particles in partially occupied Landau level can respond to arbitrarily weak electric field, thus forming compressible regions. In the fractional regime, the elementary charged excitations in the compressible regions are Laughlin quasielectrons and quasiholes.

The counterpropagating edge channels pass near the saddle points, where tunneling may occur. Thus, in the range of $B$ where the interference oscillations are observed, the filling of the edge channels is determined by the saddle point filling.[17] This allows to determine the saddle point density from the $R_{XX}(B)$ and $R_{XY}(B)$ magnetotransport, Fig. 3. The Landau level filling $\nu = hn/eB$ is proportional to the electron density $n$, accordingly the constriction $\nu$ is lower than the bulk $\nu_B$ in a given $B$. The island center $n$ is estimated to be 3% less than $n_B$ at $V_{FG} = 0$, the constriction - island center density difference is ~7%. Thus, the whole island is on the same plateau for strong quantum Hall states with wide plateaus, such as $f = 1$ and $1/3$. While $\nu$ is a variable, the quantum Hall exact filling $f$ is a quantum number defined by the quantized Hall resistance as $f = h / e^2 R_{XY}$.

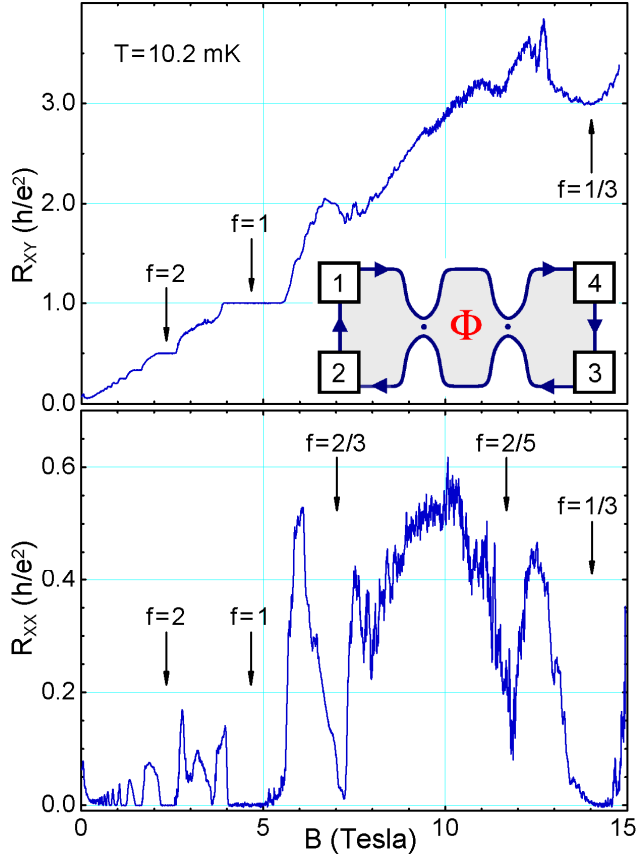

FIG. 3. The diagonal ($R_{XX}$) and Hall ($R_{XY}$) resistance of the interferometer device at zero front-gate voltage. The quantized plateaus allow to determine the filling factor in the constrictions. The fine structure is due to quantum interference effects in the residual disorder potential, including the interferometric conductance oscillations as a function of magnetic flux through the island. Inset: the chiral edge channel electron interferometer concept; dots show tunneling.

In the integer quantum Hall regime the Aharonov-Bohm ring is formed by the two counter-propagating chiral edge channels passing through the constrictions. The backscattering occurs by quantum tunneling at the saddle points in the constrictions, Fig. 1, which complete the interference path. The relevant particles are electrons of charge $-e$ and Fermi statistics, thus we can obtain an absolute calibration of the Aharonov-Bohm path area and the backgate action of the interferometer. Figure 4 shows conductance oscillations for $f = 1$; analogous oscillations were also observed for $f = 2$. The oscillatory conductance $\delta G = \delta R_{XX} / R_{XY}^2$ is calculated from the $R_{XX}$ data after subtracting a smooth background. The smooth background has two contributions: the bulk conduction at $\nu_B$ outside the bulk plateau regions, and the non-oscillatory inter-edge backscattering conductance in the interferometer. Extrapolated to $V_{FG} = 0$,[14,17] the $f = 1$ magnetic field oscillation period is $\Delta_B = 1.86$ mT. This gives the interferometer path area $S = h / e \Delta_B = 2.22 \ \mu m^2$, the radius $r = 840$ nm.

In the FQH regime, $f = 1/3$, we observe the interferometric oscillations as a function of magnetic field, Fig. 4. This is the first experimental observation of $e/3$ quasiparticle interference oscillations when the island filling is 1/3 throughout. Extrapolated to $V_{FG} = 0$, the magnetic field oscillation period is $\Delta_B = 1.93$ mT. Assuming the flux period is $\Delta_\Phi = h/e$, this gives the interferometer path area $S = h/e\,\Delta_B = 2.14\ \mu m^2$, the radius $r = 825$ nm. The island edge ring area is strictly determined by the requirement that the edge channels pass near the saddle points in the constrictions. Classically, increasing $B$ by a factor of ~3 does not affect the electron density distribution in the island at all. Quantum corrections are expected to be small for a large island containing ~3,500 electrons. Indeed, the $f = 1/3$ interferometer path area is within ±3% of the integer value, where ±3% is the estimated experimental uncertainty. The integer regime oscillations have an $h/e$ fundamental flux period, we conclude that the flux period of the 1/3 FQH oscillations is also $\Delta_\Phi = h/e$.

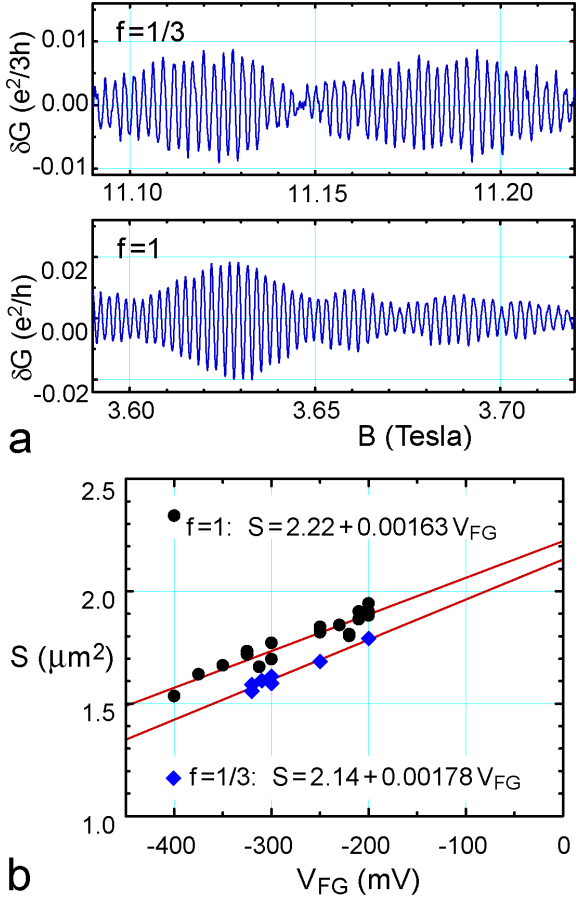

FIG. 4. (a) Representative interference conductance oscillations for electrons ($f = 1$) and for $e/3$ quasiparticles ($f = 1/3$). Moderate $V_{FG} \approx -250$ mV is applied to increase the oscillation amplitude; this reduces the island electron density and shifts the region of oscillations to lower $B$. (b) Interference path area $S = h/e\,\Delta_B$ as a function of front-gate voltage. Extrapolated to $V_{FG} = 0$, the magnetic field periods $\Delta_B$ are equal within the experimental uncertainty. Thus the flux period is $\Delta_\Phi = h/e$ in both regimes.

We use the backgate technique[6,11] to directly measure the charge period in the fractional regime. We calibrate the backgate action $\delta Q/\delta V_{BG}$, where $Q$ is the electronic charge within the Aharonov-Bohm path. The calibration is done by evaluation of the coefficient $\alpha$ using the experimental oscillation periods in

$$\Delta_Q = \alpha(\Delta_{V_{BG}}/\Delta_B),\qquad(1)$$

setting $\Delta_Q = e$ in the integer regime. Note that Eq. (1) normalizes the backgate voltage periods by the experimental $B$-periods, approximately canceling the variation in device area, for example, due to a front-gate bias. The coefficient $\alpha$ in Eq. (1) is known *a priori* in quantum antidots to a good accuracy because the antidot is completely surrounded by the quantum Hall fluid.[6,11] But, in an interferometer, the island is separated from the 2D electron plane by the front-gate etch trenches, so that its electron density is not expected to increase by precisely the same amount as $n_B$, which necessitates the calibration.

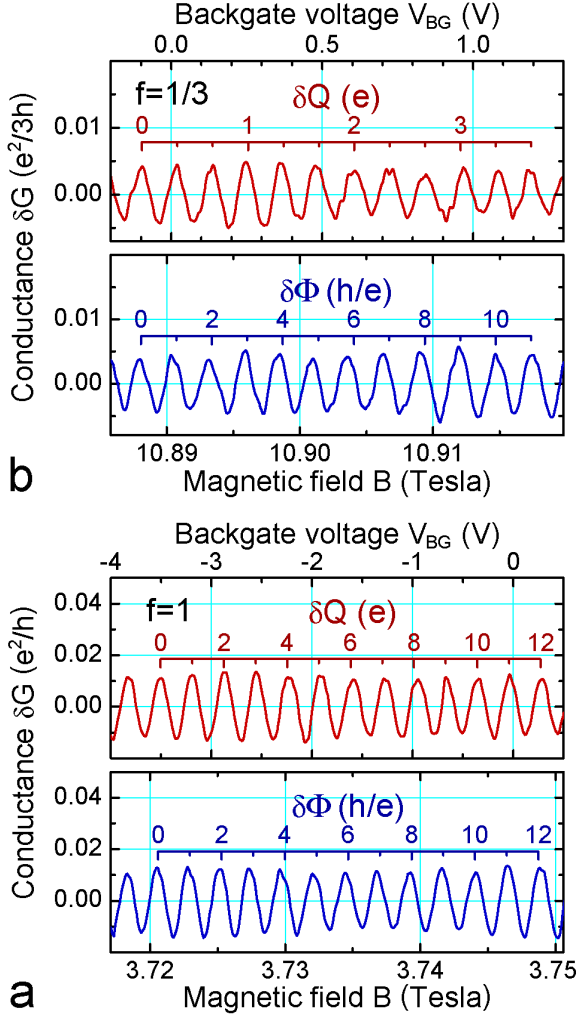

FIG. 5. Matched sets of oscillatory conductance data giving the $e/3$ charge period. (a) The interferometer device is calibrated using the conductance oscillations for electrons, $f = 1$. (b) This calibration gives the charge period for the Laughlin quasielectrons $q = (0.328 \pm 0.010)e$. The magnetic flux period $\Delta_\Phi = h/e$, the same in both regimes, implies anyonic statistics of the fractionally charged quasiparticles.

Figure 5 shows the oscillations as a function of $V_{BG}$ for $f = 1$ and $1/3$ and the analogous oscillations as a function of $B$. At each filling, the front-gate voltage is the same for the (vs $V_{BG}$, vs $B$) set. The $f = 1$ period $\Delta_{V_{BG}}$ corresponds to increment $\Delta_N = 1$ in the number of electrons within the interference path. We obtain $\Delta_{V_{BG}} = 315$ mV, $\Delta_B = 2.34$ mT, and the ratio $\Delta_{V_{BG}}/\Delta_B = 134.3$ V/T (front-gate $V_{FG} = -210$ mV for these data). This period ratio is 0.92 of that obtained in quantum antidots,[11] consistent with expectation. For the 1/3 FQH oscillations we obtain $\Delta_{V_{BG}} = 117.3$ mV, $\Delta_B = 2.66$ mT, and the ratio $\Delta_{V_{BG}}/\Delta_B = 44.1$ V/T (front-gate $V_{FG} = -315$ mV for these data). Using the integer calibration in the same device, the $e/3$ quasiparticle experimental charge period is $\Delta_Q = 0.328e$, some 1.7% less than $e/3$. To the first order, using the $\Delta_{V_{BG}}/\Delta_B$ ratio technique cancels dependence of the $V_{BG}$ and $B$ periods on the interferometer area (and $V_{FG}$ bias). The scatter of the quasiparticle charge values obtained from several matched data sets in this experimental run is $\pm 3\%$.

## III. ANALYSIS AND DISCUSSION
### A. Related work

Single-particle theory predicts flux period $\Delta_\Phi = h/q$ for charge $q$ particles.[22-24] This period is also expected for many-particle systems if the particle exchange statistics is integer (Fermi or Bose). These predictions seemed to be verified in experiments reporting conductance fluctuations in narrow-channel heterostructure samples, which reported flux periods consistent with $3h/e$ at filling 1/3.[25,26] These experiments were interpreted as fortuitously realizing a quantum antidot in the narrow channel, formed by an accidental disorder potential configuration. However, charge fluctuations consistent with $q = e$ were reported in the same samples.[26] Similar $R_{XX}$ fluctuations are seen on high-$B$ side of the $f = 2$, 1 and 1/3 plateaus in Fig. 3. We have seen such fluctuations in all constricted 2D samples over many years; however, no *periodic* fluctuations were ever detected in such regime. Subsequent experiments on narrow channels in the integer quantum Hall regime attribute such aperiodic resistance fluctuations to conductance via multiple scattering centers.[27,28]

Experiments on lithographically-defined quantum antidots have reported clear $\Delta_\Phi = h/e$ flux and $\Delta_Q = e/3$ charge periods for $f = 1/3$ plateaus.[6,11,12] In view of these results, the theoretical interpretation of the narrow-channel experiments has become: "It is our current belief that most likely their sample fortuitously realized a version of the interferometer discussed here".[16] It is stated frequently that the quantum antidot period of $h/e$ is determined by gauge invariance. Adding flux $h/e$ through the hole in the antidot (electron vacuum, where the wave-function vanishes) can be annulled by a singular gauge transformation, leaving the many-electron system in the same state as before.[29-32] However, in experiments magnetic field is varied rather than flux is inserted in the region of electron vacuum. The flux change results from change in uniform applied magnetic field so that the interacting electron system does reconstruct periodically, and the situation is likely more subtle. Several theoretical models[16,33-37] have been advanced in view of the narrow-channel experiments. The gauge invariance argument is manifestly not applicable in the interferometer geometry, where there is no electron vacuum within the interference path. A singular gauge transformation ("inserting flux") can be done only where electron wave function does not vanish, and would result in excitation or destruction of quasiparticles, the system is not the same as before.

### B. Flux and charge periods

Our experimental results can be understood as follows. The essential physics of the quasiparticle interferometer is discussed in Refs. 38, 16, 9 and 10. The experimental periods are the same as in quantum antidots, comprising addition of one quasiparticle only. When filling $v < 1/3$, as in quantum antidots, addition of flux $h/e$ to an area occupied by the 1/3 FQH condensate creates a quantized vortex, an $e/3$ quasihole. However, the interferometric oscillations are observed to occur at filling $v > 1/3$, when quasielectrons are added to the condensate. This is consistent with the principal difference between the $e/3$ interferometer and the antidots being that in quantum antidots the FQH fluid surrounds electron vacuum, while in the present interferometer the island contains the 1/3 FQH fluid everywhere within the interference path. Addition of flux reduces the number of $-e/3$ quasielectrons, the electron system is not the same as prior to flux addition, the added flux cannot be annulled by a singular gauge transformation. Another subtle difference is that a single quasihole can always be created in the 1/3 condensate, while creation of a single quasielectron is not possible in an *isolated* FQH droplet, for example.[39,40] Also, periods of $\Delta_\Phi = 3h/e$ and $\Delta_Q = e$ have been predicted in certain 1/3 FQH non-equilibrium models.[16,33,35] We therefore discuss our experimental results via quasielectron configurations in the FQH ground state, as a more restrictive case.

In an unbounded FQH fluid, changing $\nu$ away from the exact filling $f$ is accomplished by creation of quasiparticles; the ground state consists of the $\nu = f$ condensate and the matching density of quasiparticles.[3-5,36] Starting at $\nu = f$, changing magnetic field adiabatically maintains the system in thermal equilibrium. The equilibrium electron density, determined by the positively charged donors, is not affected. In present geometry, changing $B$ also changes the flux $\Phi = BS$ through the area $S$ enclosed by the interference path. In the $\nu > 1/3$ regime, decreasing $\Phi$ by $h/e$ increases the number of quasielectrons in $S$ by one, $\Delta_N = 1$, and decreases by $+e/3$ the negative FQH condensate charge within $S$. The quasielectron is created out of the 1/3 condensate, the condensate density changes by $+e/3S$, the charge within the interference path does not alter and still neutralizes the positive donor charge.

This process can be expressed in terms of the Berry phase $\gamma$ of the encircling $-e/3$ quasielectron, which includes the Aharonov-Bohm and the statistical contributions.[8,31,36] When there is only one quasiparticle of charge $q = \pm e/3$ present, its orbitals are quantized by the Aharonov-Bohm condition $\gamma_m = (|q|/\hbar)\Phi_m = 2\pi m$ to enclose flux $\Phi_m = mh/|q|$ with $m = 0,1,2,\ldots$ .[22] These quantized quasiparticle orbitals enclose $-em$ of the underlying 1/3 condensate charge. When other quasiparticles are present, in addition to the Aharonov-Bohm phase, quantization of the Berry phase includes a term describing mutual braiding statistics of the quasiparticles.[8] The total phase is quantized in increments of $2\pi$:

$$\Delta_\gamma = \frac{q}{\hbar}\Delta_\Phi + 2\pi\Theta_{1/3}\Delta_N = 2\pi, \qquad (2)$$

where $q = -e/3$ is the charge of the interfering quasielectron, and $\Theta_{1/3}$ is the statistics of the $-e/3$ quasielectrons. The first term in Eq. (2) contributes $(-e/3\hbar)(-h/e) = 2\pi/3$, the second term must contribute $4\pi/3$, giving an anyonic statistics $\Theta_{1/3} = 2/3$.

The same Berry phase equation describes the physically different process of the island charging by the backgate. Here, in a fixed $B$, increasing positive $V_{BG}$ increases the 2D electron density. The period consists of creating one $-e/3$ quasielectron out of the 1/3 condensate within the interference path, which causes the path to shrink by the area containing flux $h/e$. This is possible because the condensate is not isolated from the 2D bulk electron system, there is no Coulomb blockade, and the condensate charge within the interference path can increment by $+e/3$, any fractional charge imbalance ultimately supplied from the contacts. Thus, an $-e/3$ quasielectron is excited out of the condensate, $\Delta_N = 1$, the fixed condensate density is restored from the contacts, the interference path shrinks by area $h/eB$ so that flux $\Delta_\Phi = -h/e$ in Eq. (2), FQH fluid charge within the interference path does not neutralize the donors by $-e/3$.

## C. Experimental implications for proposed non-abelian interferometers

Interest in topological quantum computation with anyons[41-43] and in an experimental demonstration of a non-abelian system has lead to proposals of interferometric experiments.[44-48] These proposals have focused on the FQH states 5/2 and 12/5, both in the second Landau level.[49,50] The proposed experiments share the feature of an edge-channel chiral interferometer geometry where the edges are brought close by a depletion potential, thus allowing tunneling amplitudes strong enough to measure conductance oscillations. This aspect of the proposed devices makes them similar to those realized by us for abelian FQH states.

The present primary-filling interferometer is possible because the constrictions are relatively weakly depleted: the saddle point electron density is ~0.90 of the bulk density and ~0.93 of the island center density. In the earlier interferometers[9,10,13-15] and quantum antidots[6,11] the constrictions were much more depleted. The electron edge channels are supported by the gradient of the confining potential, which is more steep for the etched walls, compared to the front gate-induced depletion. Even so, in the experiments reported here we had to use a negative front-gate voltage in the FQH regime to enhance the amplitude of conductance oscillations.

This poses a major experimental obstacle to observation of an interferometric oscillations for any of the FQH states in the second Landau level, all of them occurring at fillings $3 > \nu > 2$. The typical plateau width $\Delta \nu / 2 f$ is only 0.01 in the second Landau level. In order to have only one quantum Hall filling within the interference path this necessitates that the saddle point density is 0.99 of the island center density. It is appears to be a very challenging experimental problem to be able to control depletion potential with so fine resolution, while allowing the tunneling distance to be reasonable. In addition, is not evident whether the gradient of the confining potential, at position where the electron density is 0.99 that of the island center, is sufficient to define a stable edge channel. If a larger constriction depletion is used, more than one FQH states are likely to form in the island so that identification of the origin of the interference signal, if any, may prove ambiguous. However, there is no fundamental reason why such an experiment is impossible to do, and the basic importance of the anticipated results warrants experimental exploration of this regime.

## IV. CONCLUSIONS

In conclusion, we realized a novel primary-filling $e/3$ quasiparticle interferometer where an $e/3$ quasiparticle executes a closed path around an island containing the 1/3 FQH fluid only. The central results obtained, the flux and charge periods of $\Delta_\Phi = h/e$ and $\Delta_Q = e/3$ are robust. Both the Aharonov-Bohm and the charging periods accurately correspond to excitation of one $-e/3$ quasielectron within the interference path and are interpreted as imposed by the anyonic braiding statistics $\Theta_{1/3} = 2/3$ of FQH quasiparticles.

## ACKNOWLEDGMENTS

We acknowledge discussions with D. V. Averin and B. I. Halperin. This work was supported in part by the NSF and by U.S. ARO.